\documentclass[aps,prd,preprint,superscriptaddress,tightenlines,nofootinbib]{revtex4}

\usepackage{graphicx}% Include figure files
\usepackage{dcolumn}% Align table columns on decimal point
\usepackage{bm}% bold math

\def\pbinv{pb$^{-1}$}
\def\calB{{\cal B}}

\def\ie{{\it i.e.}}

\def\Ecm{E_{cm}}
\def\DeltaM{\Delta M}

\def\pip{\pi^+}
\def\pim{\pi^-}
\def\piz{\pi^0}
\def\pipm{\pi^\pm}

\def\etaprime{\eta'}

\def\Kp{K^+}
\def\Km{K^-}
\def\Kz{K^0}

\def\KS{K^0_S}

\def\Dp{D^+}
\def\Dm{D^-}
\def\Dz{D^0}

\def\Ds{D_s}
\def\Dsp{D_s^+}
\def\Dsm{D_s^-}
\def\Dsstar{D_s^*}
\def\Dspstar{D_s^{*+}}

\def\Dspm{D_s^\pm}
\def\Dsmp{D_s^\mp}
\def\Dspmstar{D_s^{*\pm}}
\def\Dsmpstar{D_s^{*\mp}}

\def\NDDbar{N_{\Dsstar\Ds}}

\def\jbar{\bar{j}}

\def\Bi{\calB_i}
\def\Bj{\calB_j}
\def\Ni{N_i}
\def\effi{\epsilon_i}
\def\Bjbar{\calB_{\jbar}}
\def\Njbar{N_{\jbar}}
\def\effjbar{\epsilon_{\jbar}}
\def\Nijbar{N_{i\jbar}}
\def\effijbar{\epsilon_{i\jbar}}

\def\Mbc{M_{bc}}
\def\EDs{E(\Ds)}
\def\pDs{{\mathbf p}(\Ds)}
\def\MDs{M(\Ds)}
\def\MDsp{M(\Dsp)}
\def\MDsm{M(\Dsm)}
\def\Mt{M_T}

\def\MDspdg{M_{\Ds}}

\def\Fig#1{Fig.~\ref{#1}}
\def\Sec#1{Sec.~\ref{#1}}
\def\Tab#1{Table~\ref{#1}}

\catcode`\@=11 % @ signs can be used
\newcommand{\Begitem}{\begin{list}
{\csname\@itemitem\endcsname}
{\ifnum \@itemdepth >3 \@toodeep\else \advance\@itemdepth 1 
\edef\@itemitem{labelitem\romannumeral\the\@itemdepth} 
  \parsep  2pt plus 1pt minus 1pt 
  \parskip 0pt plus 1pt minus 1pt 
  \topsep  0pt plus 1pt minus 1pt             
  \itemsep 0pt plus 1pt minus 1pt \fi}}
\newcommand{\Enditem}{\end{list}}
\catcode`\@=12 % @ signs cannot be used

\begin{document}

%\preprint line(s) will be ignored for PRL/PRD
%\preprint{CLEO Draft YY-NNA} % For paper draft CBX YY-NN -> Draft YY-NNA
%\preprint{CLEO CONF YY-NN}   % For conference papers
%\preprint{ICHEP ABSnnn}      % For conference papers
\preprint{CLEO CONF 06-13}       % for CLNS notes
%\preprint{ICHEP}         % for CLNS notes

\title{\boldmath Measurement of Absolute Hadronic Branching Fractions of $\Ds$
Mesons}
\thanks{Submitted to the 33$^{\rm rd}$ International Conference on High Energy
Physics, July 26 - August 2, 2006, Moscow}

% for conference papers (ask CLEOAC for appropriate text)
%\thanks{Submitted to the 31$^{\rm st}$ International Conference on High Energy
%Physics, July 2002, Amsterdam}

%-------- INSERT HERE ------------
% Your author list goes here  REMOVE EVERYTHING to END INSERT and
% replace with your authorlist (ask cleoac).

\author{N.~E.~Adam}
\author{J.~P.~Alexander}
\author{K.~Berkelman}
\author{D.~G.~Cassel}
\author{J.~E.~Duboscq}
\author{K.~M.~Ecklund}
\author{R.~Ehrlich}
\author{L.~Fields}
\author{L.~Gibbons}
\author{R.~Gray}
\author{S.~W.~Gray}
\author{D.~L.~Hartill}
\author{B.~K.~Heltsley}
\author{D.~Hertz}
\author{C.~D.~Jones}
\author{J.~Kandaswamy}
\author{D.~L.~Kreinick}
\author{V.~E.~Kuznetsov}
\author{H.~Mahlke-Kr\"uger}
\author{P.~U.~E.~Onyisi}
\author{J.~R.~Patterson}
\author{D.~Peterson}
\author{J.~Pivarski}
\author{D.~Riley}
\author{A.~Ryd}
\author{A.~J.~Sadoff}
\author{H.~Schwarthoff}
\author{X.~Shi}
\author{S.~Stroiney}
\author{W.~M.~Sun}
\author{T.~Wilksen}
\author{M.~Weinberger}
\affiliation{Cornell University, Ithaca, New York 14853}
% ...
%\author{(CLEO Collaboration)} %FOR PRD_SPECIAL_CHANGEME
\collaboration{CLEO Collaboration} %FOR PRL,CLNS
\noaffiliation

%-------- END INSERT ------------

%please hard code the date when you have a final draft and submit to CLEOAC
\date{24 July 2006}

\begin{abstract}
We report preliminary measurements of absolute hadronic branching fractions of $\Ds$ mesons determined using a double tag technique.  These measurements are from 195 pb$^{-1}$ of $e^+e^-$ collisions recorded at center of mass energies near 4.17~GeV with the CLEO-c detector at CESR. We obtain absolute branching fractions for $\Dsp$ decays to 
$\KS\Kp$,
$\Km\Kp\pip$,
$\Km\Kp\pip\piz$,
$\pip\pip\pim$,
$\pip\eta$, and 
$\pip\etaprime$.
We discuss the problems inherent in measuring accurately the branching fraction for $\Dsp \to \phi\pip$, which is often used as a reference mode for measurement of other $\Dsp$ branching fractions, and provide a measurement of a branching fraction that may be useful for this purpose.  
\end{abstract}

\maketitle

\section{Introduction}

Absolute measurements of hadronic charm meson branching fractions play a central role in the study of the weak interaction because they serve to normalize many important $B$ and $D$ meson branching fractions.  The importance of many of these branching fractions is enhanced by their roles in determining elements of the Cabibbo-Kobayashi-Maskawa (CKM) matrix~\cite{ckm}. 

Absolute $\Ds$\footnote{Generally $\Ds$ will refer to either $\Dsp$ or $\Dsm$, and specification of an explicit $\Dspm$ state and its decay daughters will imply a corresponding relationship for the $\Dsmp$ and its daughters.} branching fractions have not been measured as accurately as those for $\Dp$ and $\Dz$ \cite{PDG}, due to difficulties in determining the total number of $\Ds$ mesons produced in an experiment.  The current best measurements of $\calB(\Dsp\to\phi\pip)$, which is usually used to normalize measurements of other $\Ds$ decay modes, are CLEO and BaBar results obtained using $B^0 \to D^{*-}\Dspstar$ events and a missing mass technique to extract the number of $\Dsp$ \cite{cleods,babards}.

In this contribution we describe measurements of absolute $\Ds$ branching fractions using a double tag technique.  This technique was pioneered by the MARK-III Collaboration~\cite{markiii-1,markiii-2} and more recently exploited in our precision measurements of absolute $\Dp$ and $\Dz$ branching fractions \cite{cleodpdz}.  This technique relies on the fact that charm or strange quarks are always produced with the corresponding antiquark ($c\bar{c}$ or $s\bar{s}$) in $e^+e^-$ annihilation.  Hence, below the thresholds for $\Dsp\Dm \bar{K}^0$ and $\Dsp\bar{D}^0 K^-$ near $\Ecm = 4.33$~GeV, production of a $\Dsp$ is always associated with production of a $\Dsm$, so observation of a $\Dspm$ can serve as a tag for production of a $\Dsmp$.  In this experiment, data were accumulated near $\Ecm = 4.17$~GeV (below the threshold for $\Dspmstar\Dsmpstar$  production) where the production of $\Dsp\Dsm$ pairs is dominated by $e^+e^-\to\Dspmstar\Dsmp$ \cite{cleoscan} with the subsequent decays $\Dspmstar\to\Dspm \gamma$ or $\Dspm\pi^0$. 

Following the MARK~III Collaboration, events in which either a $\Dsp$ or $\Dsm$ is reconstructed without reference to the other particle are called single tag (ST) events, while events in which both the $\Dsp$ and $\Dsm$ are reconstructed are called double tag (DT) events.  Absolute branching fractions can then be determined from the ratios of DT events to ST events.  If $CP$ violation is negligible, then the branching fractions $\Bj$ and $\Bjbar$ for $\Dsp \to j$ and $\Dsm\to \jbar$ are equal.  However, the efficiencies $\effi$ and $\effjbar$ for detection of these modes may be somewhat different since the cross sections for scattering of particles on nuclei may depend on the signs of the particles.  With the assumption that $\Bj = \Bjbar$, the numbers $\Ni$ and $\Njbar$ of $\Dsp \to i$ and $\Dsm\to \jbar$ ST events will be,  
\[
\Ni = \NDDbar \Bi \effi ~~\mathrm{and}~~ \Njbar = \NDDbar \Bj \effjbar,
\]
where $\NDDbar$ is the number of $\Dspmstar\Dsmp$ events produced in the
experiment.  Then, the number of DT events with $\Dsp \to i$ and $\Dsm\to \jbar$ will be,
\[ 
\Nijbar = \NDDbar \Bi\Bj \effijbar, \]
where $\effijbar$ is the efficiency for detecting the DT. 
Hence, the ratio of DT events ($\Nijbar$) to ST events
($\Njbar$) provides an absolute measurement of the branching fraction
$\Bi$,
\[ 
\Bi = {\Nijbar \over \Njbar}{\effjbar \over \effijbar}.\] 
Note that $\effijbar \approx \effi\effjbar$, so many systematic uncertainties
nearly cancel in the ratio $\effjbar/\effijbar$, and a branching fraction $\Bi$ obtained using this procedure is nearly independent of the efficiency of the tagging
mode. Of course, $\Bi$ is sensitive to $\effi$ and its uncertainties.
Finally, the number of $\Dsp\Dsm$ pairs that were produced is given by,
\[
\NDDbar = {\Ni\Njbar \over \Nijbar}\; {\effijbar \over \effi\effjbar}.
\]

We studied six decay modes,
$\Dsp\to\KS\Kp$,
$\Dsp\to\Km\Kp\pip$,
$\Dsp\to\Km\Kp\pip\piz$,
$\Dsp\to\pip\pip\pim$,
$\Dsp\to\pip\eta$, and 
$\Dsp\to\pip\etaprime$, and their charge conjugates in this analysis.  For each mode we report the average of the branching fractions for $\Dsp$ and $\Dsm$. We detected neutral kaon candidates in the $\KS \to \pi^+\pi^-$ mode, $\eta$ candidates in the $\eta\to\gamma\gamma$ mode, and $\eta'$ candidates in the $\eta' \to \pi^+\pi^-\eta$ mode.

\section{The CLEO Detector and Particle Selection}

The CLEO-c detector is a modification of the CLEO~III
detector~\cite{cleoiidetector,cleoiiidr,cleorich}, in which the silicon-strip
vertex detector was replaced with a six-layer vertex drift chamber, whose
wires are all at small stereo angles to the chamber axis~\cite{cleocyb}.  The
charged particle tracking system, consisting of this vertex drift chamber and
a 47-layer central drift chamber~\cite{cleoiiidr}, operates in a 1.0~T
magnetic field, oriented along the beam axis. The momentum resolution
of the tracking system is approximately 0.6\% at
$p=1$~GeV/$c$. Photons are detected in an electromagnetic calorimeter,
composed of 7800 CsI(Tl) crystals~\cite{cleoiidetector}, which
attains a photon energy resolution of 2.2\% at $E_\gamma=1$~GeV
and 5\% at 100~MeV. The solid angle coverage for charged and neutral
particles of the CLEO-c detector is 93\% of $4\pi$.  We utilize two
particle identification (PID) devices to separate $K^\pm$ from $\pipm$: 
the central drift chamber, which provides measurements of ionization energy
loss ($dE/dx$), and, surrounding this drift chamber, a cylindrical
ring-imaging Cherenkov (RICH) detector~\cite{cleorich}, whose active 
solid angle is 80\% of $4\pi$.  The combined
$dE/dx$-RICH PID system has a pion or kaon efficiency
$>90$\% and a probability of pions faking kaons (or vice versa) $<5$\%.

The response of the detector is studied with a detailed
GEANT-based~\cite{geant} Monte Carlo (MC) simulation of the CLEO detector
for particle trajectories generated by EvtGen~\cite{evtgen}
and final state radiation (FSR) predicted by PHOTOS~\cite{photos}.
Simulated events are processed in a fashion similar to data.
The integrated luminosity ($\cal{L}$) of the data sample is measured 
using $e^+e^-$ Bhabha events in the calorimeter~\cite{lumins}, where the event
count normalization is provided by the detector simulation.

Charged tracks are required to be well-measured and to satisfy 
criteria based on the track fit quality.  They must also be consistent with
coming from the interaction point in three dimensions. Pions and kaons are
identified by consistency with the expected $dE/dx$ and RICH information,
when available.  We select $K^0_S$ candidates from pairs of oppositely-charged tracks that are constrained to a vertex, and that have $|M(\pip\pim) - M_{\Kz}| < 6.3$~MeV$/c^2$, where $M_{\Kz}$ \cite{PDG} is the known $\KS$ mass.  The 6.3~MeV$/c^2$ limit is approximately $3\sigma$.  

We form $\pi^0$ ($\eta$) candidates from photon pairs with invariant mass within 3 standard deviations ($\sigma$) of the known $\pi^0$ ($\eta$) mass, $M_{\piz}$ ($M_{\eta}$) \cite{PDG}, where $\sigma\approx 5$--7 MeV/$c^2$ depending on photon energy and location.
These candidates are then fit kinematically with their masses
constrained to $M_{\piz}$ ($M_{\eta}$).  We reconstruct $\etaprime$ candidates in the 
decay mode $\etaprime\to\pip\pim\eta$ and require $0.9478 < M(\etaprime) < 0.9678$~GeV$/c^2$.

\section{Data Sample and Candidate Selection}

The data sample for this measurement consists of 179~\pbinv\ collected at $\Ecm = 4.17$~GeV, 10~\pbinv\ collected at $\Ecm = 4.16$~GeV, and 6~\pbinv\ collected at $\Ecm = 4.18$~GeV.  We collected the 4.16~GeV data and 4.18~GeV data in a  scan of center-of-mass energies designed to determine the optimal energy for studying $\Ds$ decays \cite{cleoscan}.  

\begin{figure}[htb]
\includegraphics[width=4in]{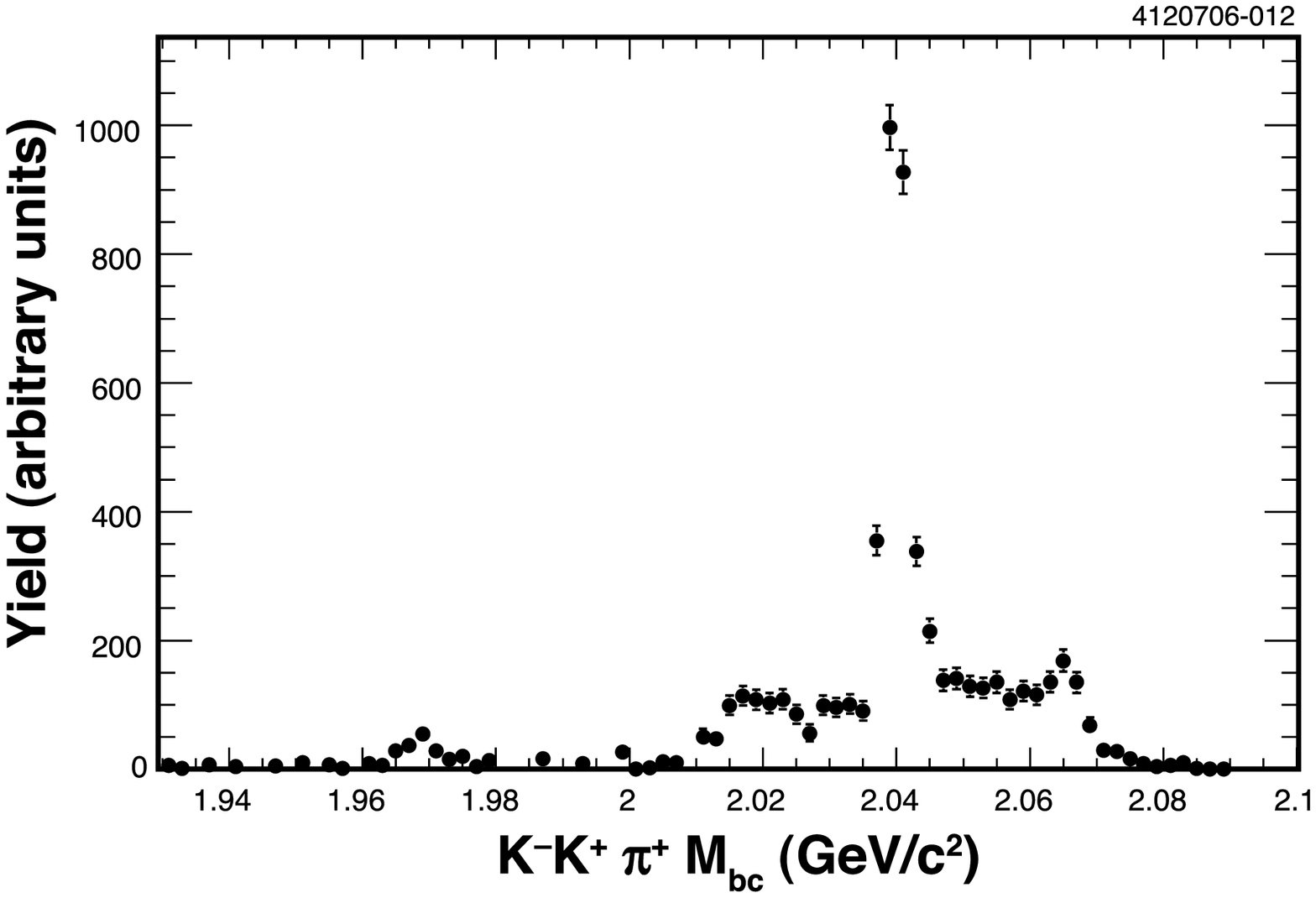}
\caption{$\Mbc$ distribution for $\Dspm \to \Km\Kp\pipm$ events in data.}
\label{fig:mbcdist}
\end{figure}

Most of the $\Ds$ mesons at $\Ecm$ near 4.17~GeV come from $\Dspmstar\Dsmp$
events, and we impose kinematic cuts on the $\Ds$ candidates to require this
topology. In such events, $\Dspmstar \to \Dspm\gamma$ or $\Dspm\piz$, so one of the $\Ds$ is produced directly and the other is a secondary from $\Dsstar$ decay.  We use two variables to select $\Ds$ decay candidates and isolate signals from backgrounds.  One variable is the candidate invariant mass, 
\begin{equation}
\MDs c^2 = \sqrt{\EDs^2 - \pDs^2c^2}, \label{eq:invmass}
\end{equation}
where $\pDs$ and $\EDs$ are the total momentum and energy, respectively, of the daughters of the candidate $\Ds$.  The particle energies are computed from particle momenta using the masses determined by PID.  The other variable is the beam constrained mass, 
\begin{equation}
\Mbc c^2 \equiv \sqrt{E_0^2 - {\pDs}^2c^2} \label{eq:mbc}
\end{equation}
obtained by substituting the beam energy  $E_0$ for the energy $\EDs$ of the candidate.  
We make the following observations and choices based on these observations for this analysis:
\Begitem 
\item We do not require full reconstruction of the event; \ie, we make no attempt to find the $\gamma$ or $\pi^0$ resulting from the $\Dsstar$ decay.  This avoids a large efficiency loss and avoids substantial contributions to systematic errors that would arise from soft photon detection and which would affect ST and DT yields differently.
\item The value of $\Ecm$ determines the momenta of the $\Ds$ and $\Dsstar$, so the momentum of the direct $\Ds$ is tightly constrained, and the momentum of the $\Ds$ from $\Dsstar$ decay has a somewhat wider distribution about the same value. 
\item The $\Mbc$ distribution of direct $\Ds$ mesons has a peak about 0.010~GeV$/c^2$ wide centered at 2.04~GeV$/c^2$ -- somewhat above $\MDspdg = 1.9683$~GeV/$c^2$ \cite{PDG}.  The $\Mbc$ distribution of the $\Ds$ daughters from $\Dsstar$ decay is broader and relatively flat from about 2.015~GeV$/c^2$ to about 2.065~GeV$/c^2$.  These two components of the $\Mbc$ mass distribution are clearly evident in \Fig{fig:mbcdist}.  The upper limit of the $\Mbc$ distribution is just the beam energy $E_0$, while the lower limit is determined by the maximum momentum of the secondary $\Ds$ candidates. 
\item  In our selection of $\Ds$ candidates, we use $\Mbc$ as a proxy for this momentum, since it is insensitive to variations in $\Ecm$ and allows easy integration of data taken at different energies.  
\item We use the invariant mass $\MDs$ of the $\Ds$ candidate as our primary analysis variable.  It has very little correlation with $\Mbc$ in the region of interest.
\Enditem

We require $|\MDs - \MDspdg| < 0.085$~GeV$/c^2$ and a ``loose'' cut $\Mbc > 2.01$~GeV$/c^2$, for $\Ds$ candidates in all decay modes.  For the $\Dsp\to\Km\Kp\pip\piz$ and $\Dsp\to\pip\pip\pim$ modes, we require additional ``tight'' cuts of $\Mbc > 2.034$~GeV$/c^2$ and $\Mbc > 2.035$~GeV$/c^2$, respectively.  The loose cut includes the entire range of $\Mbc$ resulting in maximum efficiency.  The tight cut selects all of the directly-produced $\Ds$ and over half of the indirect $\Ds$ as well.  This reduces the background while reducing the impact of initial-state radiation smearing of efficiency.
\medskip

\noindent Furthermore, for ST candidates we require:
\smallskip
\Begitem
\item $\Dsp\to\KS\Kp$: We reject candidates with $M(\Kp\pim) > 1.83$~GeV$/c^2$ to eliminate a small background from $\bar{D}^0 \to \Kp\pim$ decays.
\item $\Dsp\to\Km\Kp\pip$: We reject candidates with $1.845 < M(\Km\Kp) < 1.88$~GeV$/c^2$ to eliminate a small background from $\Dz \to \Km\Kp$ decays.
\item $\Dsp \to \Km\Kp\pip\piz$: We require $\mathbf{p}_{\piz} > 0.1$~GeV$/c$ to reduce a background from $D^{*\pm}$ and $D^{*0}$ decays, and we reject candidates with $1.86 < M(\Km\Kp\pip) < 1.88$~GeV$/c^2$ to eliminate a small background from $\Dp\to\Km\Kp\pip$ decays.
\item $\Dsp\to\pip\pip\pim$: We reject candidates if either $\pip\pim$ candidate has $475 < M(\pip\pim) < 520$~MeV$/c^2$ (to eliminate $\KS\to\pip\pim$ decays), or either candidate has $1.840 < M(\pip\pim) < 1.885$~GeV$/c^2$ (to eliminate $\Dz\to\pip\pim$ decays).  Also, when the $\pim$ is assumed to be a $\Km$ and is assigned the $\Km$ mass, we reject the $\Dsp$ candidate if either $\Km\pip$ combination had $1.845 < M(\Km\pip) < 1.880$~GeV$/c^2$ to eliminate background from $\Dz\to\Km\pip$.
\item In every event, for each mode and each charge, we choose the best ST candidate.  We accept the candidate with the smallest value of $|\Mbc - \widehat{M}|$, where $\widehat{M} = 2.040553$~GeV$/c^2$ is the value at which $\Mbc$ peaks for daughters of direct $\Ds$ decays. An event can have candidates in more than one mode for a given $\Ds$ charge. 
\Enditem 

\noindent Furthermore, for DT candidates we require:
\smallskip
\Begitem
\item In a DT event, one of the $\Ds$ will be direct and the other will be the daughter of a $\Dsstar$ decay.  We impose the loose $\Mbc$ requirement on both of the $\Ds$ candidates.
\item Each of the $\Ds$ candidates in a DT event must pass the ST  requirements described above.  
\item To resolve multiple candidates, we select the candidate with total invariant mass closest to $2\MDspdg$.  This variable is orthogonal to the difference between the masses of the two candidates, which is used in selecting signal and background sideband events as described below.
\Enditem 

\section{Efficiencies and Yields}

To develop our analysis procedures and estimate efficiencies, we simulated two types of $\Ds$ meson decays:
\Begitem
\item ``signal'' decays in which a $\Dsp$ or $\Dsm$ decays in one of the 6 modes measured in this analysis, and 
\item ``generic'' decays in which a $\Dsp$ or $\Dsm$ decays in a manner
  consistent with the decay modes and branching fractions in the PDG
  compilation \cite{PDG}, extended with additional decays that are expected to
  be present.
\Enditem
Using these two types of simulated decays, we generated three types of Monte
  Carlo (MC) events:
\Begitem
\item generic Monte Carlo events, in which both the $\Dsp$ and the
$\Dsm$ decay generically,
\item single tag signal Monte Carlo events, in which either the
$\Dsp$ or the $\Dsm$ always decays in one of the 6 modes measured in this
analysis while the $\Dsm$ or $\Dsp$, respectively, decays generically, and
\item double tag signal Monte Carlo events, in which both the $\Dsp$
and the $\Dsm$ decay in one of the 6 modes in this analysis.  
\Enditem
In these events, both the $\Dsp$ and $\Dsm$ have a 50\% probability of being
produced from a $D_s^{\pm *}$.

We also generated MC samples of continuum $u$, $d$, $s$ quark production and
production of non-strange charmed meson states, which form the major
background to this analysis. The relative production rates are set by
information obtained from the scan of this energy region \cite{cleoscan}.

We obtained the ST efficiencies and yields by fitting $\MDs$ distribution for
each mode in single tag MC events.  The fit functions and their parameters, as
well as efficiencies were determined from $\MDs$ distributions from signal
Monte Carlo simulations.%  

\begin{figure}[htb]
\includegraphics[width=3.5in]{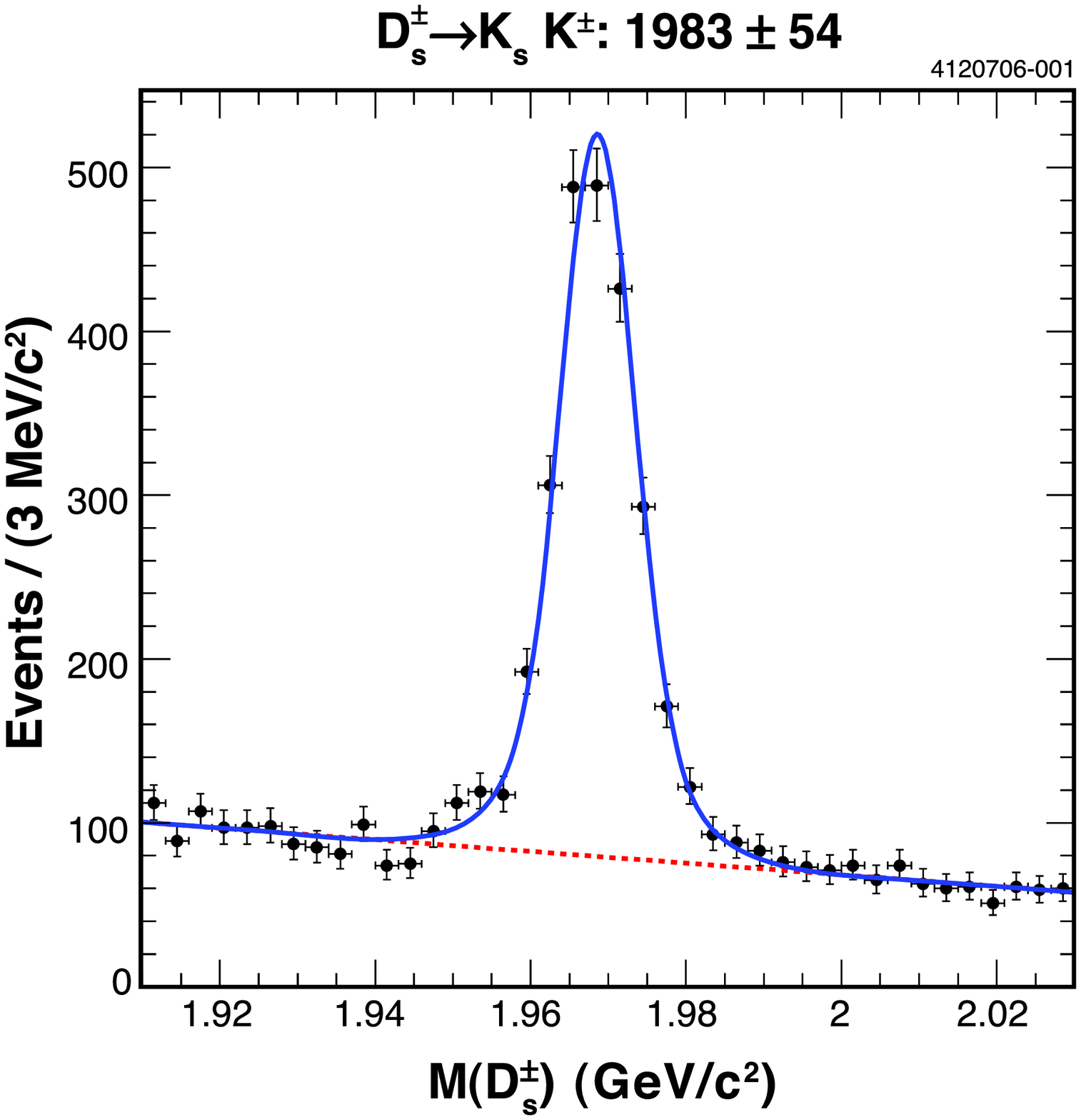}
\caption{Fits to single tag data for $\Dspm\to\KS K^\pm$.}
\label{fig:kskp}
\end{figure}

\begin{figure}[htb]
\includegraphics[width=3.5in]{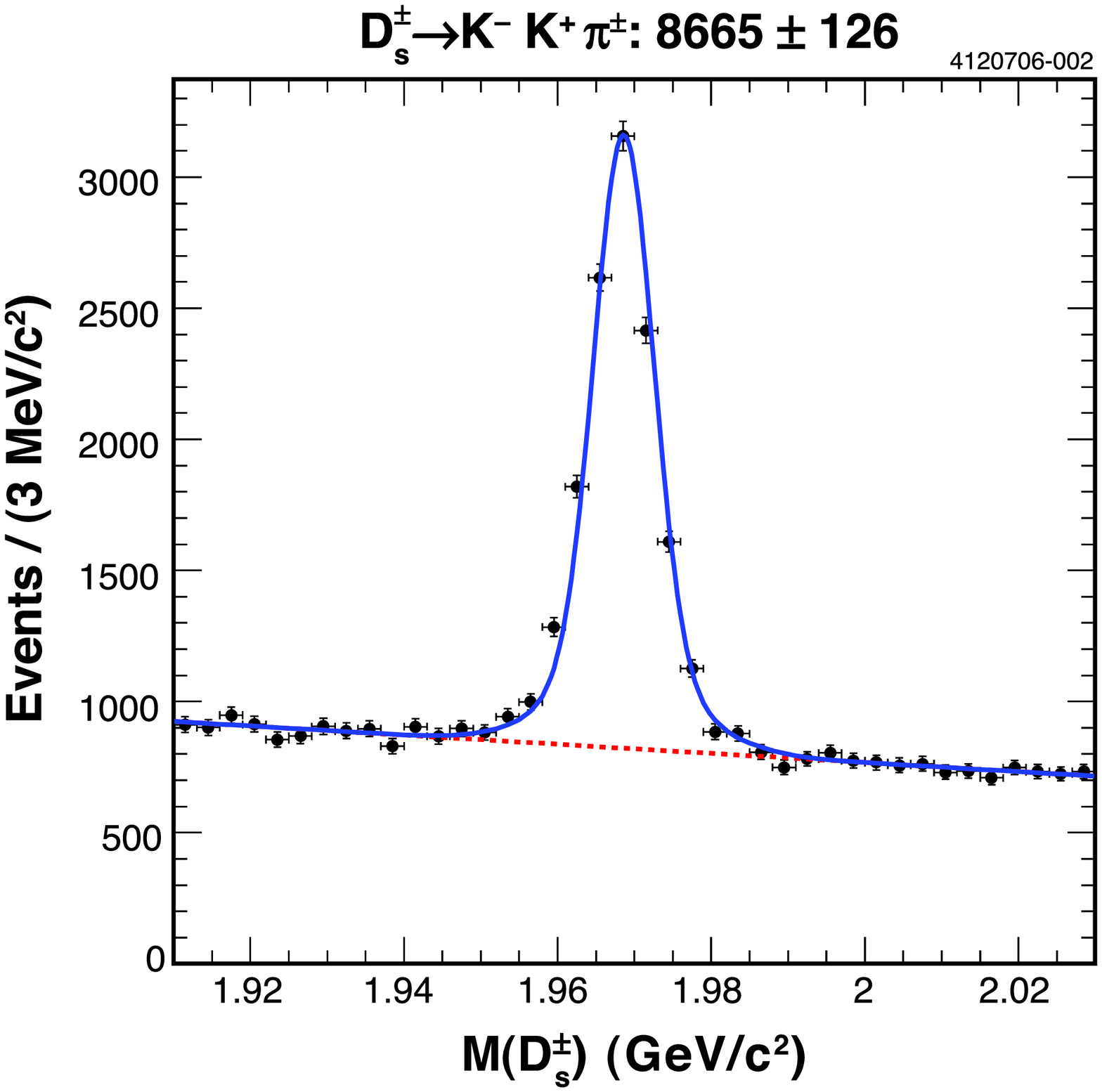}
\caption{Fits to single tag data for $\Dspm\to\Km\Kp \pipm$.}
\label{fig:kmkppip}
\end{figure}

\begin{figure}[htb]
\includegraphics[width=3.5in]{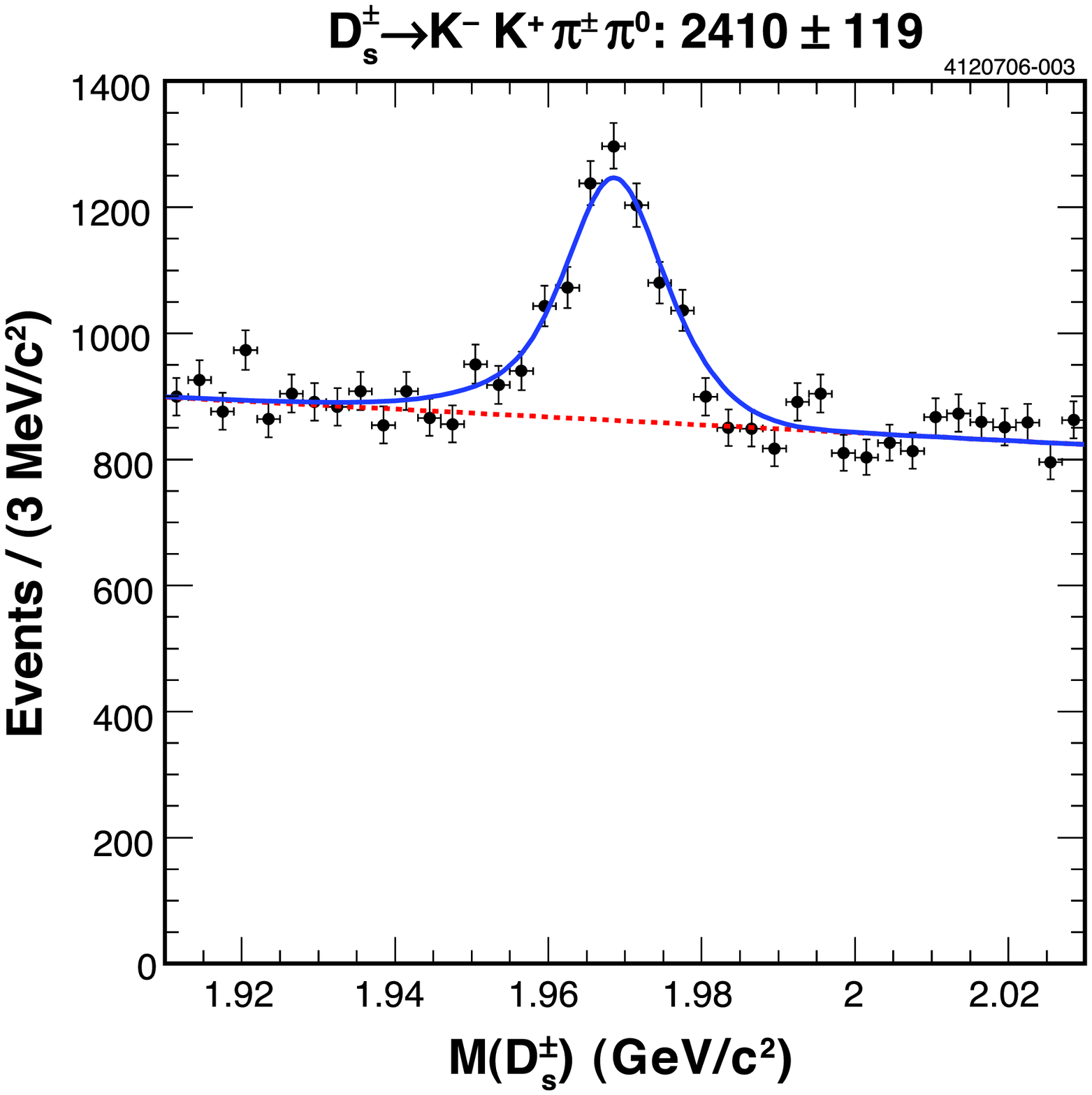}
\caption{Fits to single tag data for $\Dspm\to\Km\Kp \pipm\piz$.}
\label{fig:kmkppippiz}
\end{figure}

\begin{figure}[htb]
\includegraphics[width=3.5in]{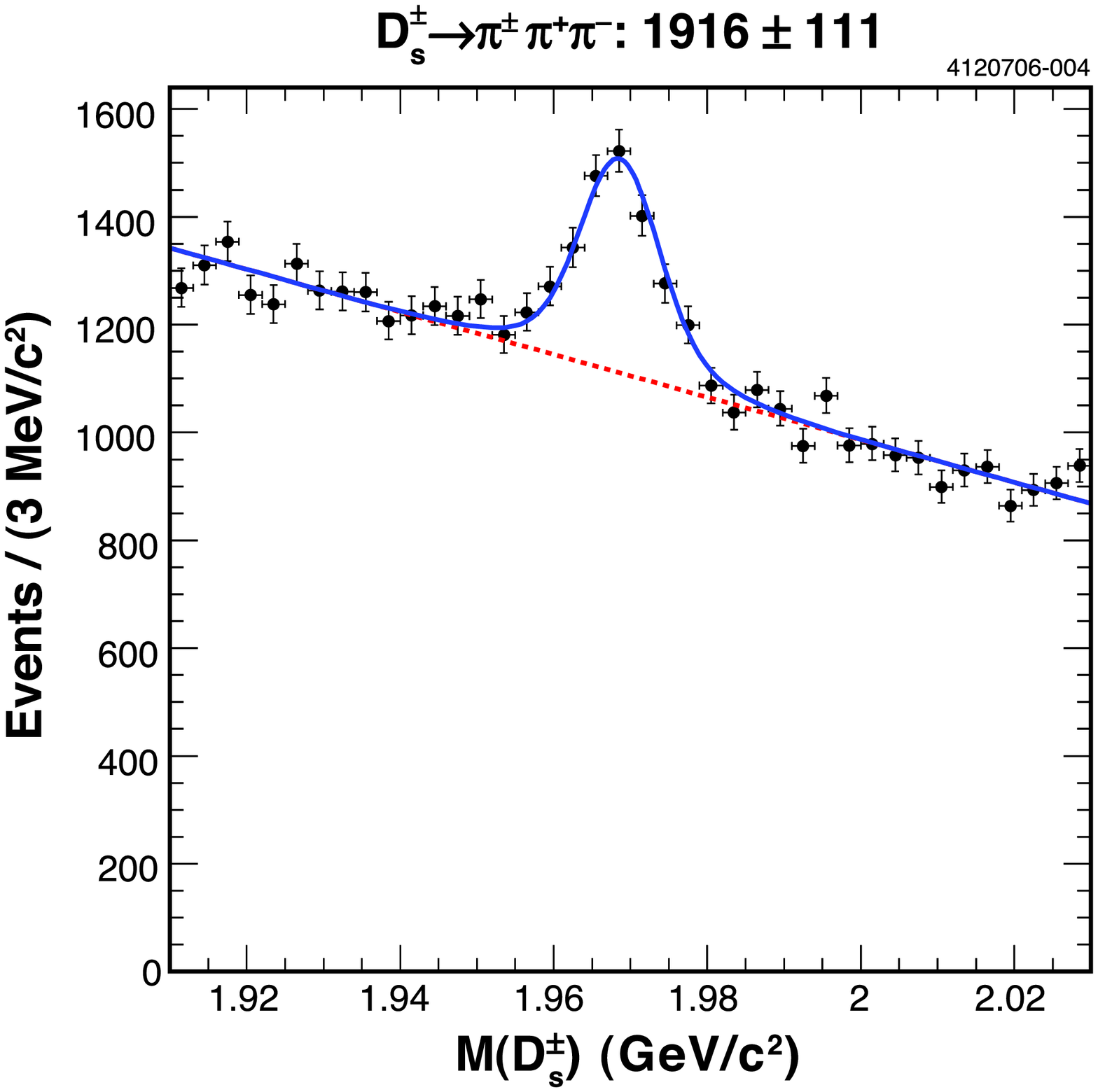}
\caption{Fits to single tag data for $\Dspm\to\pip\pim\pipm$.}
\label{fig:pippippim}
\end{figure}

\begin{figure}[htb]
\includegraphics[width=3.5in]{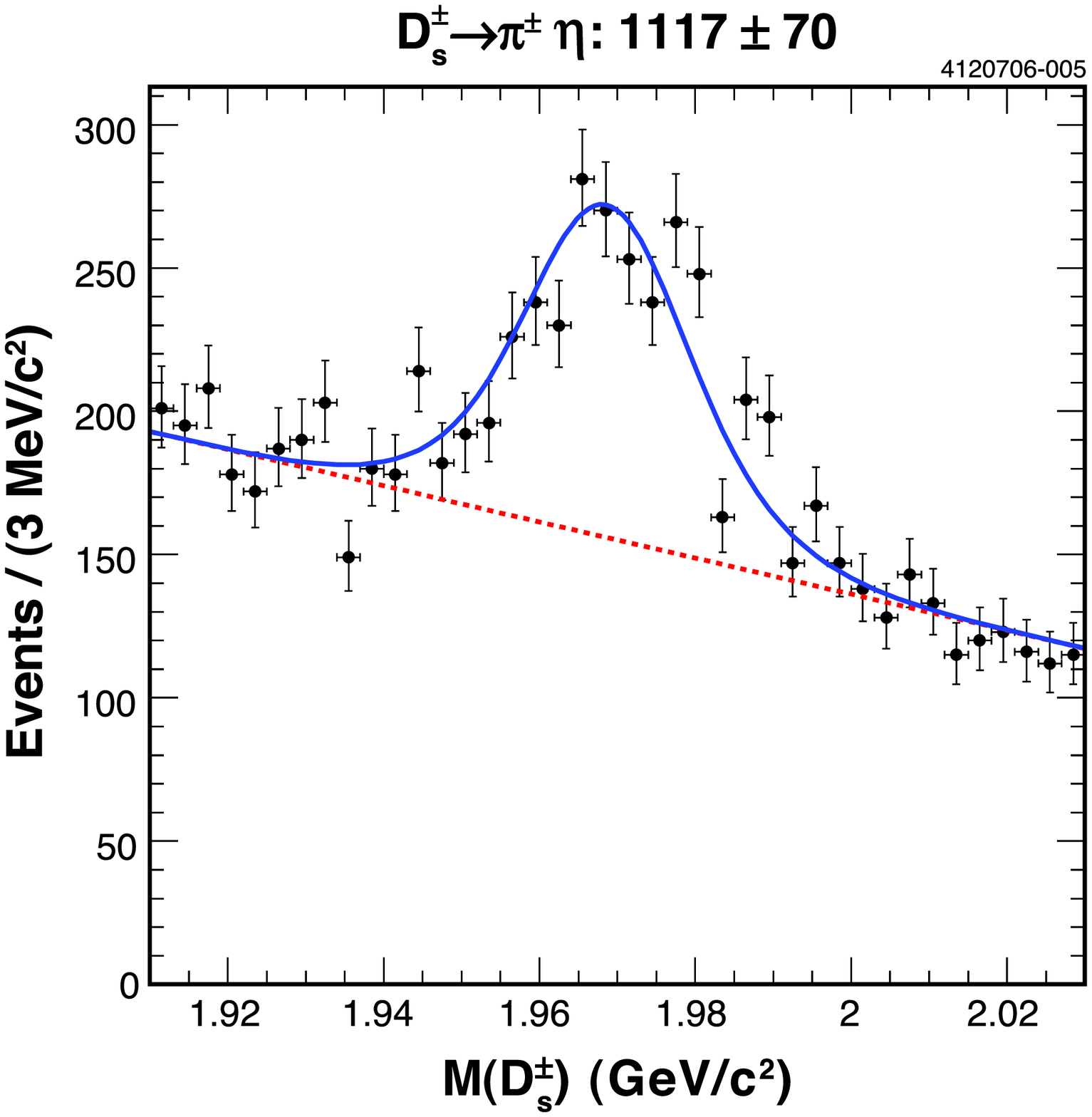}
\caption{Fits to single tag data for $\Dspm\to\pipm\eta$.}
\label{fig:pipeta}
\end{figure}

\begin{figure}[htb]
\includegraphics[width=3.5in]{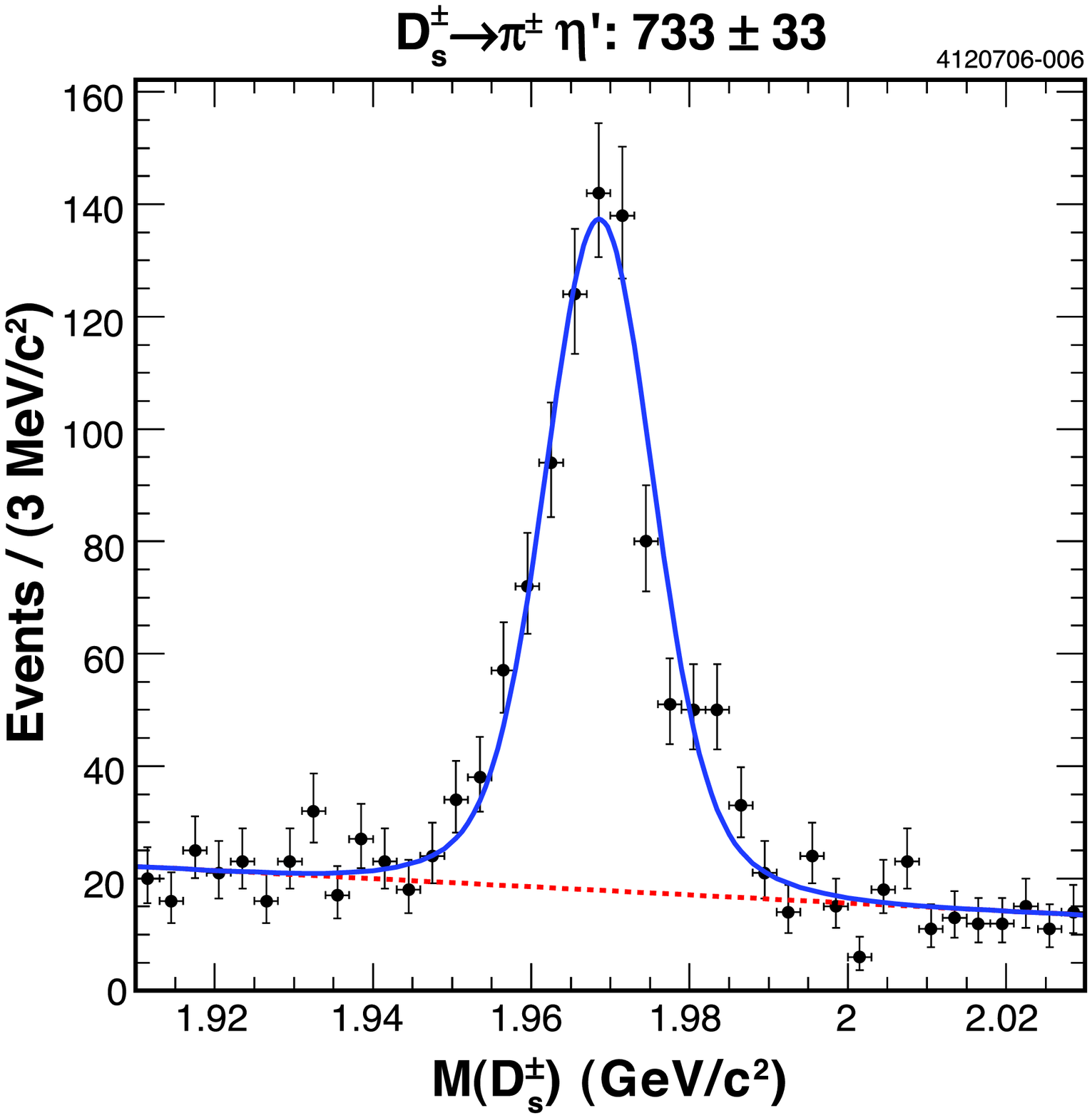}
\caption{Fits to single tag data for $\Dspm\to\pipm\eta'$.}
\label{fig:pipetaprime}
\end{figure}

The ST distributions are fit with a linear backgrounds and a signal lineshapes.  Two Gaussian distributions were chosen for the lineshapes for the  $\KS\Kp$, $\Km\Kp\pip$, $\pip\pip\pim$, and $\pip\etaprime$ modes.  The lineshapes chosen for  
$\Km\Kp\pip\piz$ and $\pip\eta$ decays was the sum of a narrow Gaussian and a wider Crystal Ball function \cite{cbfun}, whose low mass tail simulates energy leakage in the CsI calorimeter.  The lineshape parameters are fixed separately for each charge from candidates that are matched to the generator-level $\Ds$.  The ST efficiencies obtained from signal MC events are given in \Tab{tbl:steffs}.  The fits to $\MDs$ distributions in data are illustrated in Figs.~\ref{fig:kskp}--\ref{fig:pipetaprime}, and the corresponding yields from data are given in \Tab{tbl:datastyields}.
  
\renewcommand{\arraystretch}{1.1}

\begin{table}[htb]
\begin{center}
\caption{\label{tbl:steffs}Single tag efficiencies (in \%) obtained from signal MC events.  \Sec{sec:systematics} describes small corrections that must be applied to these efficiencies.}
\smallskip
\begin{tabular}{lcc}
\hline\hline
$\Dsp$ Mode & ~~~~~~~~~$\Dsp$~~~~~~~~~ & ~~~~~~~~~$\Dsm$~~~~~~~~~\\
\hline
$K_{S} K^+$ & 37.58 $\pm$ 0.24 & 37.43 $\pm$ 0.24\\
$K^- K^+ \pi^+$ & 44.79 $\pm$ 0.25 & 43.90 $\pm$ 0.25\\
$K^- K^+ \pi^+ \pi^0$ & 12.32 $\pm$ 0.19 & 12.69 $\pm$ 0.19\\
$\pi^+ \pi^+ \pi^-$ & 49.52 $\pm$ 0.26 & 49.36 $\pm$ 0.26\\
$\pi^+ \eta$ & 19.99 $\pm$ 0.21 & 19.09 $\pm$ 0.21\\
$\pi^+ \eta'$ & ~5.29 $\pm$ 0.12 & ~5.53 $\pm$ 0.12\\
\hline\hline
\end{tabular}
\end{center}
\end{table}

\begin{table}[htb]
\begin{center}
\caption{\label{tbl:datastyields}Single tag yields obtained from fits to the $\MDs$ distributions of data.}
%\smallskip
\begin{tabular}{lcccccc}
\hline\hline
& $K_{S} K^+$ & $K^- K^+ \pi^+$ & $K^- K^+ \pi^+ \pi^0$ & $\pi^+ \pi^+ \pi^-$ & $\pi^+ \eta$ & $\pi^+ \eta'$ \\
\hline
$\Dsp$ &~ 1055.0 $\pm$ 39.4 &~ 4319.1 $\pm$ 88.8 &~ 1342.3 $\pm$ 101.9 &~ 969.5 $\pm$ 79.3 &~ 547.0 $\pm$ 49.6 &~ 361.8 $\pm$ 23.4 \\
$\Dsm$ &~ ~927.6 $\pm$ 37.3 &~ 4352.2 $\pm$ 88.7 &~ 1371.9 $\pm$ 102.9 &~ 946.7 $\pm$ 77.6 &~ 570.0 $\pm$ 49.8 &~ 371.5 $\pm$ 23.6 \\
\hline\hline
\end{tabular}
\end{center}
\end{table}

We obtain double tag yields by counting the numbers of events in signal regions in $\MDsm$ vs $\MDsp$ planes and subtracting the numbers of events in sideband regions of the corresponding planes.  We find that combinatoric backgrounds from $\Dp$, $\Dz$, and continuum events tend to be distributed along lines of constant total mass, 
$\Mt \equiv \MDsp + \MDsm$.  Both signal and sideband regions use subsets of the $\Mt$ region, $3.915 < \Mt < 3.960$~GeV$/c^2$.  Within this $\Mt$ region, the signal region is $|\DeltaM| < 30$~MeV$/c^2$, while the sideband regions are $50 < |\DeltaM| < 150$~MeV$/c^2$, where $\DeltaM = \MDsp-\MDsm$.  We see no evidence for any background to peak in the signal region, so we are confident that the events in these sideband regions provide reasonable estimates of the backgrounds in the signal regions.   DT efficiencies obtained from signal MC events are given in \Tab{tbl:dteffs} and DT yields obtained from data are given in \Tab{tbl:datadtyields}. 

\begin{table}[htb]
\begin{center}
\caption{\label{tbl:dteffs}Double tag efficiencies (in \%) obtained from signal MC events.  \Sec{sec:systematics} describes small corrections that must be applied to these efficiencies.}
%\smallskip
\hspace*{-1.5em}\begin{tabular}{l|cccccc}
\hline\hline
& $K_{S} K^-$  & $K^+ K^- \pi^-$  & $K^+ K^- \pi^- \pi^0$  & $\pi^- \pi^- \pi^+$  & $\pi^- \eta$  & $\pi^- \eta'$  \\
\hline
$K_{S} K^+$  &~ 14.02 $\pm$ 0.17  &~ 17.30 $\pm$ 0.19  &~ ~6.74 $\pm$ 0.13  &~ 22.84 $\pm$ 0.21  &~ ~6.64 $\pm$ 0.12  &~ ~1.96 $\pm$ 0.07  \\
$K^- K^+ \pi^+$  &~ 16.75 $\pm$ 0.19  &~ 20.35 $\pm$ 0.20  &~ 7.65 $\pm$ 0.13  &~ 27.91 $\pm$ 0.22  &~ ~8.01 $\pm$ 0.14  &~ ~2.39 $\pm$ 0.08  \\
$K^- K^+ \pi^+ \pi^0$~  &~ ~6.74 $\pm$ 0.13  &~ ~7.88 $\pm$ 0.13  &~ ~2.40 $\pm$ 0.08  &~ 11.09 $\pm$ 0.16  &~ ~3.17 $\pm$ 0.09  &~ ~1.00 $\pm$ 0.05  \\
$\pi^+ \pi^+ \pi^-$  &~ 22.48 $\pm$ 0.21  &~ 27.69 $\pm$ 0.22  &~ 11.22 $\pm$ 0.16  &~ 35.94 $\pm$ 0.24  &~ 10.91 $\pm$ 0.16  &~ ~3.17 $\pm$ 0.09  \\
$\pi^+ \eta$  &~ ~6.68 $\pm$ 0.12  &~ ~8.34 $\pm$ 0.14  &~ ~3.21 $\pm$ 0.09  &~ 10.88 $\pm$ 0.16  &~ ~3.18 $\pm$ 0.09  &~ ~0.92 $\pm$ 0.05  \\
$\pi^+ \eta'$  &~ ~1.94 $\pm$ 0.07  &~ ~2.47 $\pm$ 0.08  &~ ~0.94 $\pm$ 0.05  &~ ~3.32 $\pm$ 0.09  &~ ~0.92 $\pm$ 0.05  &~ ~0.32 $\pm$ 0.03  \\
\hline\hline
\end{tabular}
\end{center}
\end{table}

\begin{table}
\begin{center}
\caption{\label{tbl:datadtyields}Double tag yields obtained from data by subtracting backgrounds estimated from sidebands.}
\smallskip
\begin{tabular}{lcccccc}
\hline\hline
& ~~$K_{S} K^-$~~  & ~~$K^+ K^- \pi^-$~~  & ~~$K^+ K^- \pi^- \pi^0$~~  & ~~$\pi^- \pi^- \pi^+$~~  & ~~$\pi^- \eta$~~  & ~~$\pi^- \eta'$~~  \\
\hline
$K_{S} K^+$  & 7.7  & 27.0  & 18.7  & 7.3  & 4.0  & 5.0  \\
$K^- K^+ \pi^+$  & 18.0  & 104.7  & 43.7  & 30.7  & 12.0  & 8.0  \\
$K^- K^+ \pi^+ \pi^0$  & 8.7  & 35.7  & 14.0  & 13.3  & 1.0  & 5.7  \\
$\pi^+ \pi^+ \pi^-$  & 3.3  & 22.7  & 16.0  & 13.3  & 4.7  & 4.0  \\
$\pi^+ \eta$  & 0.0  & 10.0  & 2.7  & 6.0  & 1.0  & 1.7  \\
$\pi^+ \eta'$  & 3.0  & 10.0  & 3.0  & 3.7  & 1.0  & 0.0  \\
\hline\hline
\end{tabular}
\end{center}
\end{table}

\section{Branching Fractions}

We are unable to use the standard branching fraction fitter \cite{brfit} created for our analysis of $\Dp$ and $\Dz$ branching fractions \cite{cleodpdz}.  This fitter uses $\chi^2$ minimization, which is not appropriate for the small yields and backgrounds in the DT modes.  For this analysis, we have developed a binned-likelihood hybrid fitter which utilizes Gaussian statistics for the ST modes and  Poisson statistics for the DT modes.  The likelihood is maximized with the number of $\Dspmstar\Dsmp$ mesons ($\NDDbar$) and the branching fractions for the individual modes ($\Bi$) averaged over $\Dsp$ and $\Dsm$ as the free parameters.  The hybrid fitter treats properly the correlated systematic errors among quantities affecting ST efficiencies only.  However, unlike the standard fitter, this fitter does not handle correlated systematic errors in efficiencies among double tag modes or between single tag modes and double tag modes.  On the other hand, the approximate cancellation of systematic errors between DT and ST efficiencies is still utilized in computing overall systematic errors. 

We tested this hybrid fitter extensively in MC simulations and with data from our $\Dp$ and $\Dz$ analysis.  This fitter produces results very close to the standard fitter in that analysis.  We also tested this hybrid fitter with toy MC  samples with ST and DT yields comparable to those in this analysis.  We find that the hybrid fitter shows less bias and produces pull distributions that are more nearly Gaussian and with standard deviations closer to 1.

The branching fractions and statistical errors resulting from applying the hybrid fitter to the ST and DT yields are given in \Tab{tbl:bresults}.  The table also includes the systematic errors described in the next section.
\begin{table}[htb]
\begin{center}
\caption{\label{tbl:bresults}Branching fractions obtained from the ST and DT yields, using the hybrid fitter.  The errors are statistical and systematic, respectively.}
\smallskip
\begin{tabular}{lc}
\hline\hline
$\Dsp$ Mode & $\calB$ (\%) \\
\hline
$K_{S} K^+$ & $1.50 \pm 0.09 \pm 0.05$ \\
$K^- K^+ \pi^+$ & $5.57 \pm 0.30 \pm 0.19$\\
$K^- K^+ \pi^+ \pi^0$ ~~~~& $5.62 \pm 0.33 \pm 0.51$\\
$\pi^+ \pi^+ \pi^-$ & $1.12 \pm 0.08 \pm 0.05$ \\
$\pi^+ \eta$ & $1.47 \pm 0.12 \pm 0.14$ \\
$\pi^+ \eta'$ & $4.02 \pm 0.27 \pm 0.30$\\
\hline\hline
\end{tabular}
\end{center}
\end{table}

\section{Systematic Errors}
\label{sec:systematics}

\subsection{Detector Simulation}

Using a missing mass technique, we measure efficiencies for
reconstructing tracks, $K^0_S$ decays, and $\pi^0$ decays in both
data and simulated events.  We fully reconstruct $\psi(3770)\to D\bar D$,
$\psi(2S)\to J/\psi\pi^+\pi^-$, and $\psi(2S)\to J/\psi\pi^0\pi^0$
events, leaving out one particle, for which we wish to determine
the efficiency.  The missing mass of this combination peaks at
the mass of the omitted particle, whether or not it is detected.
Then, the desired efficiency is the fraction of this peak with this
particle correctly reconstructed.  We find good agreement between efficiencies in data and simulated events.  The relative uncertainties in these determinations, 0.35\% per charged particle track,
1.1\% per $K^0_S$, and 5.0\% per $\piz$ or $\eta$.  Extrapolation from the relatively low energies of the pions in $\psi(2S)\to J/\psi\pi^0\pi^0$ decay to the higher energies encountered in our $\Ds$ decay modes contributes significantly to the larger uncertainties for $\piz$ and $\eta$ simulation.  

\subsection{Charged Particle Identification}

We study the simulation of the PID efficiencies for charged pions and kaons using decays with unambiguous particle content, such as $D^0\to K^0_S\pi^+\pi^-$ and $D^+\to K^-\pi^+\pi^+$.  We find small momentum dependent corrections to efficiencies for pions and kaons.  The uncertainties in these corrections are estimated to be less than 0.2\% for pions and 0.3\% for kaons.  We use these corrections to compute efficiency corrections which depend on the decay mode, and on submodes for modes with multiple resonant states.   The correction factors and their uncertainties are listed in \Tab{tbl:pidfix}.
The quoted uncertainties for the $\Km\Kp\pip$ and $\Km\Kp\pip\piz$ modes
cover the range of correction factors estimated for the submodes.    

\begin{table}
\begin{center}
\caption{\label{tbl:pidfix}Corrections to the efficiencies from signal MC simulations, for particle ID, resonant substructure, and their product.  The ST efficiencies given in \Tab{tbl:steffs} are corrected by the product factors in this table, while the DT efficiencies given in \Tab{tbl:dteffs} are corrected by the product factors for the two modes in the DT.  The uncertainties are the contributions of these factors to the systematic error.  The quoted uncertainties for the $\Km\Kp\pip$ and $\Km\Kp\pip\piz$ modes cover the ranges of corrections for the submodes.}
\smallskip
\begin{tabular}{lccc}
\hline\hline
Mode & PID  & Resonant     & Product \\
     &      & Substructure & \\
\hline
$\KS\Kp$         &~ 0.$989 \pm 0.003$ ~&~  1 ~&~ 0.$989\pm 0.003$ \\
$\Km\Kp\pip$     &~ $0.971 \pm 0.008$ ~&~ $1.000 \pm 0.015$ ~&~
                    $0.971 \pm 0.017$ \\
$\Km\Kp\pip\piz$ &~ $0.961 \pm 0.014$ ~&~ $1.033 \pm 0.060$ ~&~
                    $0.993 \pm 0.062$ \\
$\pip\pip\pim$   &~ $0.989 \pm 0.006$ ~&~ $1.000 \pm 0.020$ ~&~
                    $0.989 \pm 0.021$ \\
$\pip\eta$       &~ $0.998 \pm 0.002$ ~&~  1 ~&~ $0.998 \pm 0.002$\\
$\pip\etaprime$  &~ $0.968 \pm 0.006$ ~&~  1 ~&~ $0.968 \pm 0.006$\\
\hline\hline
\end{tabular}
\end{center}
\end{table}

\subsection{Resonant Substructure}

The resonant substructures of the inclusive multibody modes in this analysis are poorly understood.  Since final-state momenta and angular distributions are sensitive to the intermediate resonances in these decays, the efficiencies will depend on the resonant substructure.  We estimate possible corrections to efficiencies and the associated systematic uncertainties by estimating the number of events in our data samples in the different resonant modes.  We then estimate the range of possible corrections from the corrections for these different resonant modes.  This correction does not apply to the two-body modes 
$\KS\Kp$, $\pip\eta$, and $\pip\etaprime$.  The estimated corrections to the efficiencies and the associated systematic uncertainties for the multibody modes are included in \Tab{tbl:pidfix}. 

\subsection{Single Tag Fitting Functions}

To study possible differences between the MC lineshape and the lineshape in data, we perform two checks:  We allow the overall mass and width of the signal lineshape to float (keeping the relative normalization and ratio of widths of the two Gaussians fixed), and we allow a quadratic term in the background.  The
maximum excursions of the yields of two additional fits from the yields of the normal fits are taken to be the systematic uncertainties.  These excursions and the systematic uncertainties are included in \Tab{tbl:lineshapesyst}.

\begin{table}
\begin{center}
\caption{\label{tbl:lineshapesyst}Systematic uncertainties due to uncertainties in the ST fitting functions.  The middle two columns are yield excursions (averaged over charge) for using a quadratic background or allowing the $\MDs$ width and mass to float.  The rightmost column is the systematic uncertainty assigned to single tag yields in this mode.  All numbers are in percent (\%).}
\smallskip
\begin{tabular}{lccc}
\hline\hline
Mode & ~~Second-order~~ & Float & ~~Systematic~~ \\
     & Polynomial   &  ~~Width \& Mass~~ & Uncertainty \\
\hline
     $\KS\Kp$ &~ $-0.1$ & $+0.1$ & $~\pm$0.1\\
$\Km\Kp\pip$ &~ $-0.5$ & $+2.0$ & $\pm$2.0\\
$\Km\Kp\pip\piz$ & $+8.7$ & $-10.8$ & $\pm$10.8\\
$\pip\pip\pim$ & $+1.4$ &~ $-2.1$ &~ $\pm$2.1\\
$\pip\eta$ &~ $-8.8$ & $+11.1$ & $\pm$11.1\\
$\pip\etaprime$~ &~ $-4.8$ &~ $+4.5$ &~ $\pm$4.8\\
\hline\hline
\end{tabular}
\end{center}
\end{table}

\subsection{Initial State Radiation}

Initial state radiation (ISR) can impact the $\Mbc$ requirement used in selection of decay modes by reducing the momenta of the $\Ds$ candidates, leading to high-side tails on the $\Mbc$ distributions for primary and secondary $\Ds$ mesons.  By definition, $\Mbc$ is limited by the beam energy $E_0$

 For the loose $\Mbc$ requirement, we accept all events from the lower
 kinematic limit to the beam energy (\ie\ momenta from the upper kinematic
 limit to zero), so the cut is fully efficient regardless of ISR.  The tight cuts used for the  $\Kp\Km\pip\piz$ and $\pip\pip\pim$ modes eliminate most of the lower half of the $\Mbc$ distribution for secondary $\Ds$ from $\Dsstar$ decay. Since the effect of ISR on the $\Mbc$ distribution is independent of decay mode, we study the comparison between MC and data using the high-statistics $\Dsp\to\Kp\Km\pip$ mode.  First, we compare loose and tight yields in data and MC simulation without ISR.  We find $R_{Data} \equiv y_{L}/y_{T} = 0.807 \pm 0.008$, where $y_{L}$ and $y_{T}$ are the yields in data from the loose and tight cuts respectively.  The corresponding number from MC simulation without ISR is $R_{\overline{ISR}} = 0.814 \pm 0.003$, so the two ratios agree within 1\%.  We then study this ratio in a MC simulation with ISR, and find $R_{ISR} = 0.806 \pm 0.005$, agreeing with the other two ratios within 1\%.  We do not correct for this effect, but we include a 1\% systematic uncertainty in the ST efficiencies for the  $\Kp\Km\pip\piz$ and $\pip\pip\pim$ modes where we apply this tight $\Mbc$ cut.    

\subsection{Multiple Candidates}
\label{sec:multiple}

If there is more than one candidate for a given mode, and we select only one, there is some inefficiency associated with making the incorrect choice for this mode.  This inefficiency is modeled in MC simulations, but if the multiple candidate rates are significantly different in data and MC simulations, there is likely to be an efficiency difference.  We examine signal and sideband multiplicity distributions; the invariant mass regions for these two data samples are $1.955 < \MDs < 1.980$~GeV$/c^2$ and $1.90 < \MDs < 1.93$~GeV$/c^2$, respectively.  We compare multiplicity distributions for data and MC simulations  after subtracting sideband distributions from signal distributions.   There is good agreement between data and MC multiplicity distributions, although data multiplicities are generally marginally larger.  

If we assume we can do no better than randomly choose candidates (so, for example, the efficiency is down 50\% for two candidates, 67\% for three), we can limit the effect of this difference on the total efficiency.  We find that this effect on ST efficiencies is negligible for the $\KS\Kp$, $\Kp\Km\pip$, and $\pip\etaprime$ modes.  For the other three modes we assign systematic uncertainties to the ST efficiencies of:  3\% for $\Kp\Km\pip\piz$,  1\% for $\pip\pip\pim$, and  0.5\% for $\pip\eta$.  

\subsection{Other Possible Systematic Uncertainties}

Several potential systematic errors do not contribute significantly to the systematic uncertainties of these measurements.

We checked our analysis procedure, the internal consistency of our Monte Carlo simulations, and the utility of the hybrid fitter by analyzing a generic MC sample corresponding to approximately six times the luminosity of the current data sample.  This sample included appropriately weighted samples of all open charm channels at 4.17~GeV.  There is no significant evidence for bias in the analysis or fitting procedure, so we do not include a systematic error for our procedures.

Data were accumulated in three separate data sets; we do not observe significant variations of yields per unit luminosity over these data sets.  

The ST yields for $\Dsp$ and $\Dsm$ are consistent, so we do not include a contribution to the systematic error for the differences in efficiencies or yields for $\Dsp$ and $\Dsm$.

The $\Dsp$ daughters from the isospin violating decay $\Dspstar\to\Dsp\piz$ have  a different momentum distribution from those from $\Dspstar\to\Dsp\gamma$ decay.   
The yields of the $\Kp\Km\pip\piz$ and $\pip\pip\pim$ modes are sensitive to this momentum difference due to the tight $\Mbc$ requirements for these modes.
However, the effect is small due to the small branching fraction for the
$\Dsp\piz$ mode.  We estimate that the possible effect is of ${\cal O}
(0.14\%)$ and we ignore it.

Potential differences in $\Mbc$ resolution in MC simulations and data can
affect the efficiency for the tight cuts used for the $\Kp\Km\pip\piz$ and
$\pip\pip\pim$ modes.  The $\Mbc$ resolution $\sigma_{\Mbc}$ is dominated by
the beam energy spread. If we vary this by 10\%, we expect an efficiency
change of ${\cal O}(0.2\%)$ or less.

Because we use invariant mass as our fit variable, the only backgrounds that we expect to peak in our signal regions of $\MDs$ are those that produce the same final state daughters.  The Cabibbo suppressed decay $\Dsp\to\KS\pip$ can fake $\Dsp\to\pip\pip\pim$ decays.  As described above, we eliminate this background with cuts on the $\pip\pim$ mass for either pair, so we do not include a correction or systematic error for this possible background.  Via misidentification of the $\Kp$, the decay $\Dsp\to\Kp\pip\pim$ can fake the $\KS\pip$ mode.  We search for such a signal in data and find none, so we do not include a correction or systematic error for this potential background.  

\subsection{Summary of Systematic Errors}

\Tab{tbl:syserrors} summarizes the systematic uncertainties for the inclusive branching fractions measured in this analysis.

\begin{table}[htb]
\begin{center}
\caption{Summary of systematic uncertainties in efficiencies for inclusive branching fraction measurements.  This list excludes systematics from MC statistics and fit function systematics obtained in the data fits.  Where ranges are given, the values for individual modes can be found in \Tab{tbl:pidfix} for Particle Identification,  \Tab{tbl:lineshapesyst} for ST Fit Functions, and \Sec{sec:multiple} for Multiple Candidate Rates.}\label{tbl:syserrors}
\smallskip
\begin{tabular}{lcc}
\hline\hline
Source & Uncertainty (\%) & {Affects} \\
\hline
Detector Simulation & 0.35 & Tracks \\
& 1.1 & $\KS$   \\
& 5.0 & $\piz$  \\
& 5.0 & $\eta$  \\
Particle Identification & 0.3--1.4 & Per mode   \\
Resonant Substructure & 1.5 & $\Kp\Km\pip$  \\
 & 6.0 & $\Km\Kp\pip\piz$   \\
 & 2.0 & $\pip\pip\pim$  \\ \hline
ST Fit Functions & 0.1--11.1  & Per mode  \\
Initial State Radiation & 1.0 & $\Km\Kp\pip\piz$  \\
 & 1.0 &  ~~~$\pip\pip\pim$~~ \hfill   \\
Multiple Candidate Rates & 0.0--3.0 & Per mode \\
\hline\hline
\end{tabular}
\end{center}
\end{table}

The systematic uncertainties discussed in this section were added in quadrature to obtain the systematic uncertainties given in \Tab{tbl:bresults} for the six inclusive modes included in this analysis.  

\section{Partial Branching Fractions for \boldmath{$\Dsp \to \Km\Kp\pi^+$}}

The decay $\Dsp \to \phi\pi^+ \to \Km\Kp\pip$, being one of the largest and
 easiest to reconstruct $\Ds$ decays, has historically been used as a
 reference mode.  Many analyses take advantage of the small $\phi$ width and
 impose cuts on the $\Km\Kp$ system in a $\Km\Kp\pip$ event to enhance
 the signal significance, generally in the form of an invariant mass cut
 around the $\phi$ and
 possibly a helicity angle cut (to take advantage of the $\phi$ polarization).
Our inclusive branching fraction results do not directly connect to analyses
that select this submode for background suppression.  

\begin{figure}[htb]
\includegraphics[width=4in]{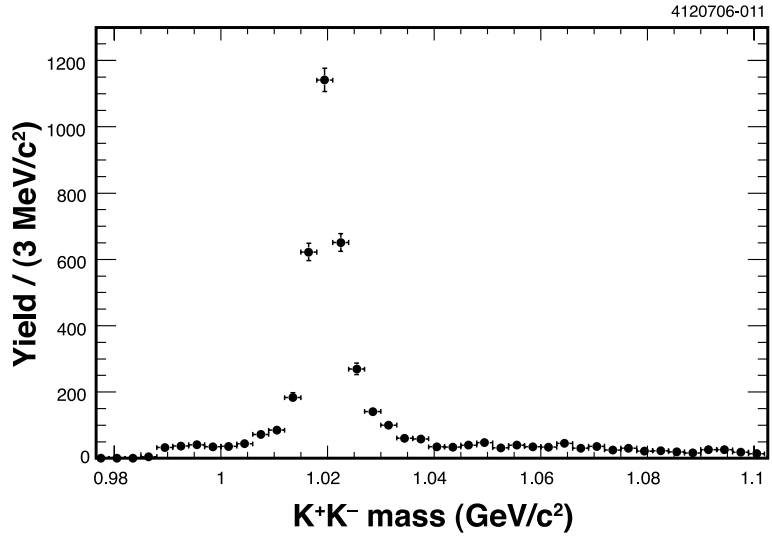}
\caption{$M(\Kp\Km)$ distribution for $\Dsp \to \Km\Kp\pip$ events with
  $M(\Kp\Km)$ near the mass of the $\phi$.}
\label{fig:mkk}
\end{figure}

We have investigated the $\Km\Kp$ invariant mass spectrum in $\Km\Kp\pip$
decays.  The mass distribution, obtained by fitting for the $\Ds$ signal in
bins of $M(\Km\Kp)$, is shown in \Fig{fig:mkk}.  Intrinsic detector resolution
on $M(\Km\Kp)$ near the $\phi$ mass is about 1 MeV$/c^2$.  We observe a clear peak
arising from the $\phi\pip$ intermediate state.  However there is a broad
signal from another source, extending from $\Km\Kp$ threshold to masses above
$M_\phi$.  This signal is found to have kinematic distributions indicative of
a scalar.  The E687 collaboration has published \cite{e687} and the FOCUS
collaboration has reported \cite{focus}
Dalitz plot analyses of $\Dsp \to \Km\Kp\pip$ and find
significant scalar contributions from $f_0(980)$ (or $a_0(980)$) in this mass
region.  Because of this extra signal, an exclusive $\phi\pip$ branching
fraction is not necessarily useful for normalization purposes, and ignoring
the scalar contribution leads to the branching fraction being ill-defined by
$\mathcal{O}(5\%)$ depending on the specific choice of cuts.

We report here measurements of the inclusive branching fraction for
$\Km\Kp\pip$ decays where the $\Km\Kp$ system lies in specific mass ranges.
We define
\[ \mathcal{B}_{\Delta M} \equiv \mathcal{B}(\Dsp\to\Km\Kp\pip) ~\textrm{
  for }~
|M(\Km\Kp) - 1.0195\textrm{ GeV}/c^2| < \Delta M\textrm{ MeV}/c^2 \]
and measure $\mathcal{B}_{10}$ and $\mathcal{B}_{20}$.

To measure these partial branching fractions we perform fits to $\Km\Kp\pip$
signal events where $M(\Km\Kp)$ passes the respective cuts to obtain yields
$N_{\Delta M}$.  The efficiencies
$\epsilon_{\Delta M}$ for event reconstruction are obtained from signal MC samples of
pure $\phi\pip$ and $f_0(980) \pip$ production.  The uncertainty arises
largely from differences in the reconstruction efficiencies for scalar and
vector angular distributions.  Uncertainties on the fit functions and
detector resolution smearing of the mass distribution contribute 0.4\%
to the systematics on the yields for both branching fractions.  We obtain the
partial branching fractions from
\[ \mathcal{B}_{\Delta M} = \frac{N_{\Delta M}}{2  \epsilon_{\Delta M} \NDDbar
 } \]
utilizing the value of $\NDDbar$ we obtain in the fit to the six
 inclusive modes.
\begin{table}
\begin{center}
\caption{\label{tbl:partialbfs}Efficiency, yields in data, and partial
  branching fraction results for two choices of $M(\Km\Kp)$ cuts in
  $\Km\Kp\pip$ events.}
\smallskip
\begin{tabular}{cccc}
\hline\hline
Branching Fraction~~ & ~~Efficiency (\%)~~ & ~~~~~~Data Yield~~~~~~ & ~~~~Result (\%)~~~~ \\
\hline
$\mathcal{B}_{10}$ & 43.3 $\pm$ 1.3 & ~3062 $\pm$ 58 $\pm$ 11~ & 1.98 $\pm$ 0.12 $\pm$ 0.09\\
$\mathcal{B}_{20}$ & 43.4 $\pm$ 1.3 & 3482 $\pm$ 63 $\pm$ 13 & 2.25 $\pm$ 0.13 $\pm$ 0.12\\
\hline\hline
\end{tabular}
\end{center}
\end{table}

The efficiencies, data yields, and partial branching fraction results are
shown in \Tab{tbl:partialbfs}.

\section{Results and Conclusions}
We have reported preliminary absolute measurements of the branching fractions
of $\Ds$ mesons to six final states, with total relative uncertainties ranging
from 6.4\% to 12.5\%.  We have also obtained measurements of two partial
branching fractions, to $\Km\Kp\pip$ substates with restricted values of $M(\Km\Kp)$.

We gratefully acknowledge the effort of the CESR staff 
in providing us with excellent luminosity and running conditions.
This work was supported by the National Science Foundation
and the U.S. Department of Energy.

\begin{figure}[htb]
\includegraphics[width=3.2in]{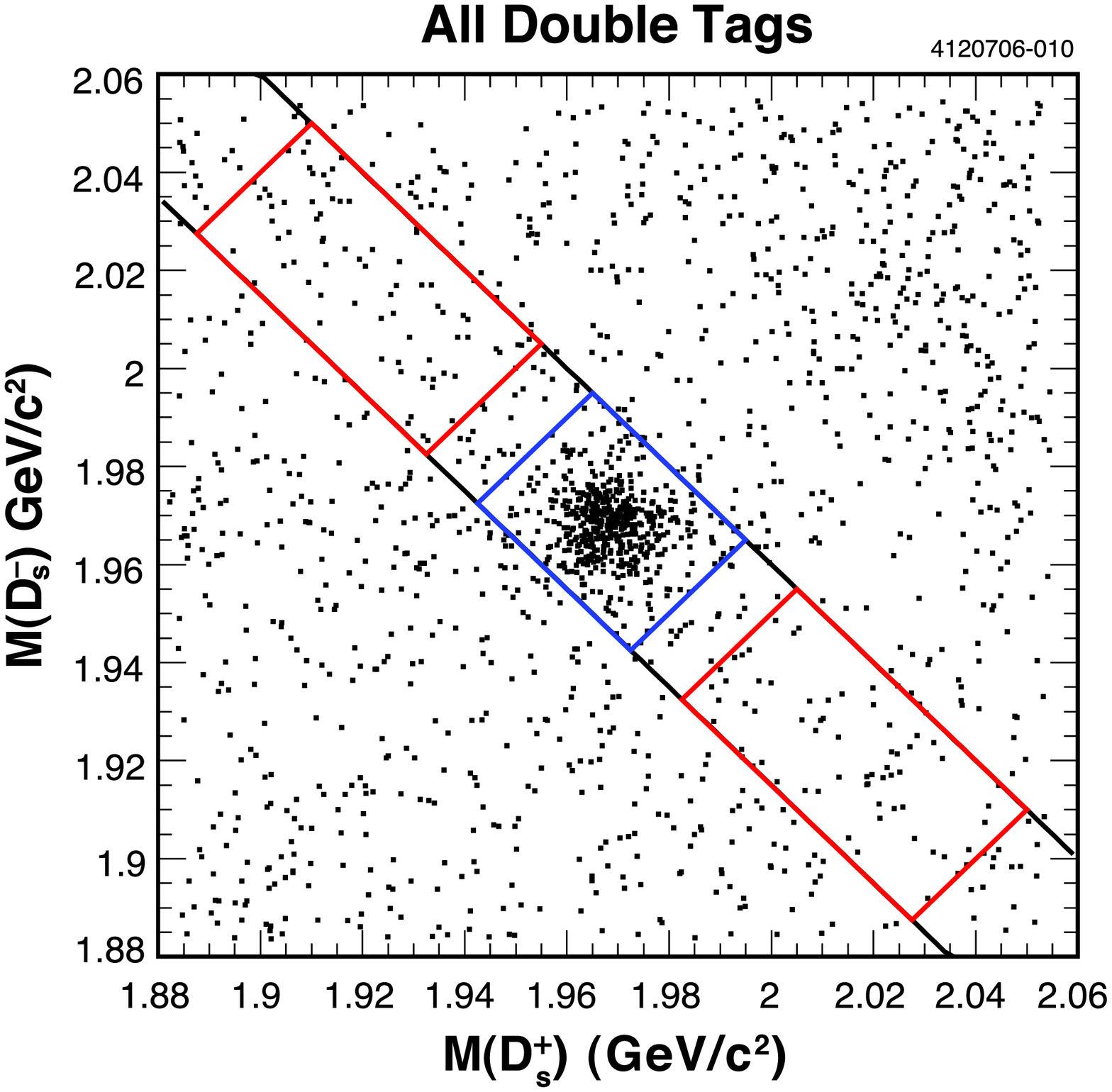}\hfill
\includegraphics[width=3.2in]{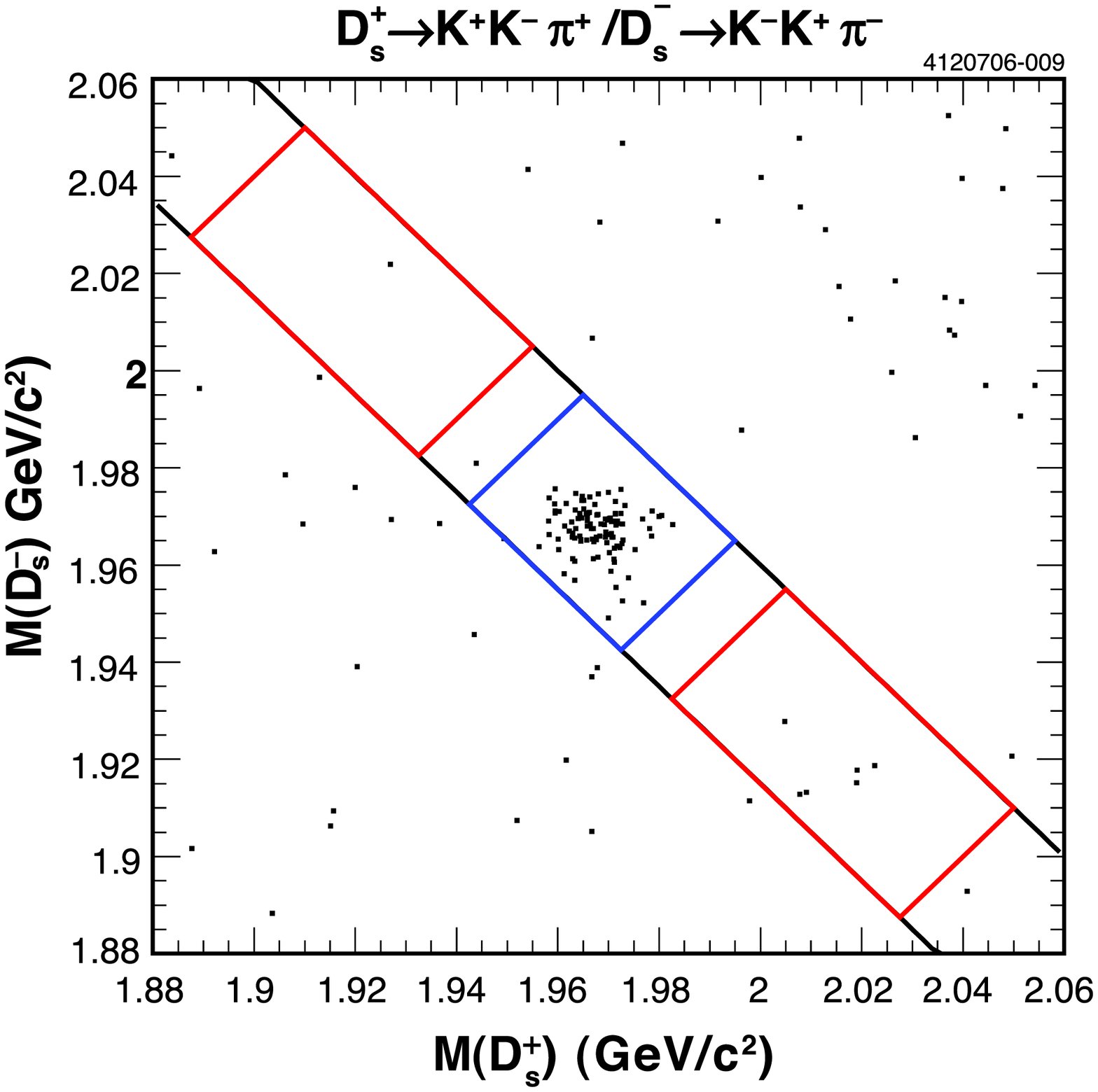}
\caption{Scatter plots of $\MDsm$ vs $\MDsp$ for (left) all six $\Ds$ decay modes and (right) $\Dspm\to\Km\Kp\pipm$.}
\label{fig:dtscatter}
\end{figure}

\begin{figure}[htb]
\includegraphics[width=3.2in]{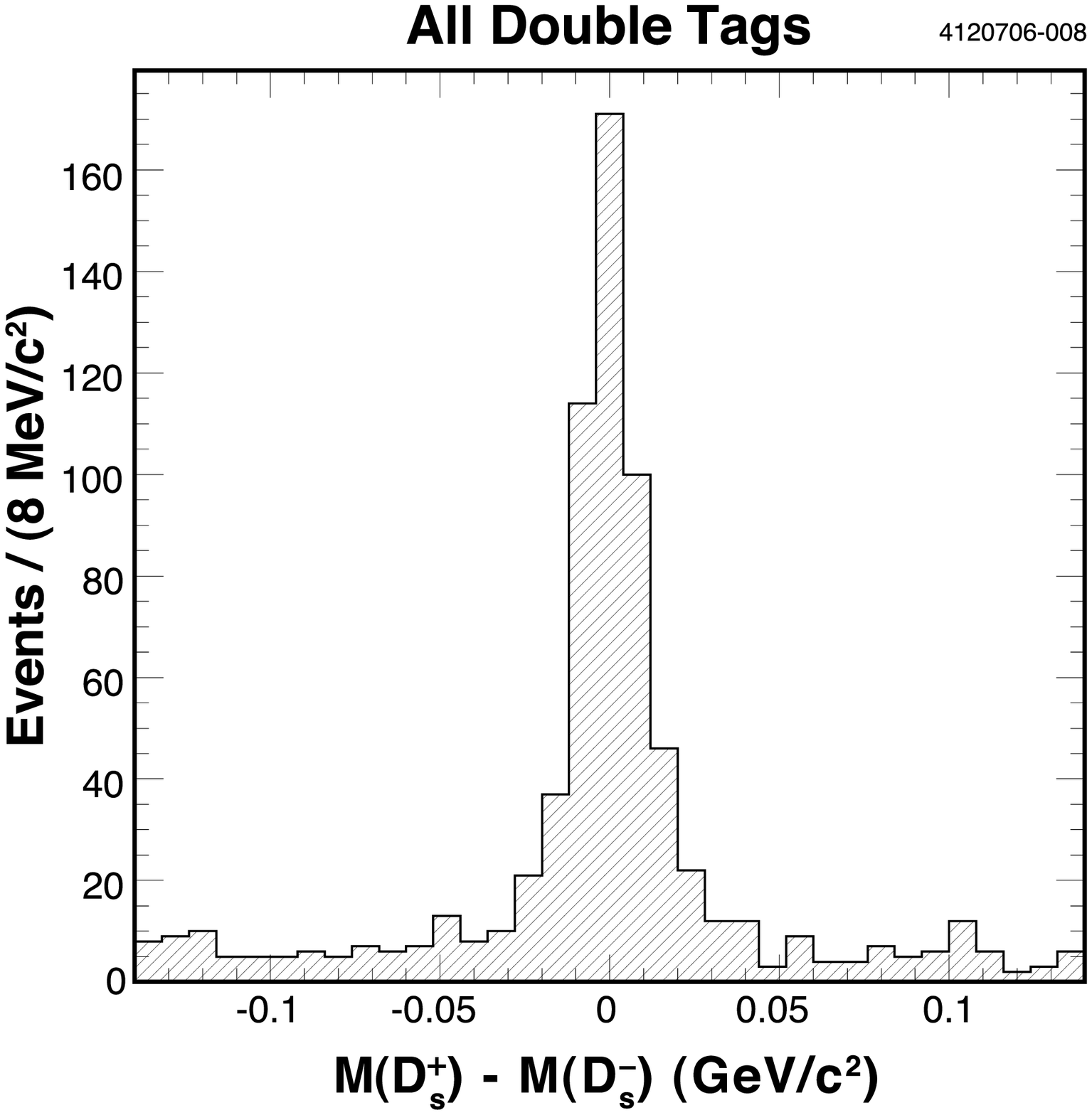}\hfill
\includegraphics[width=3.2in]{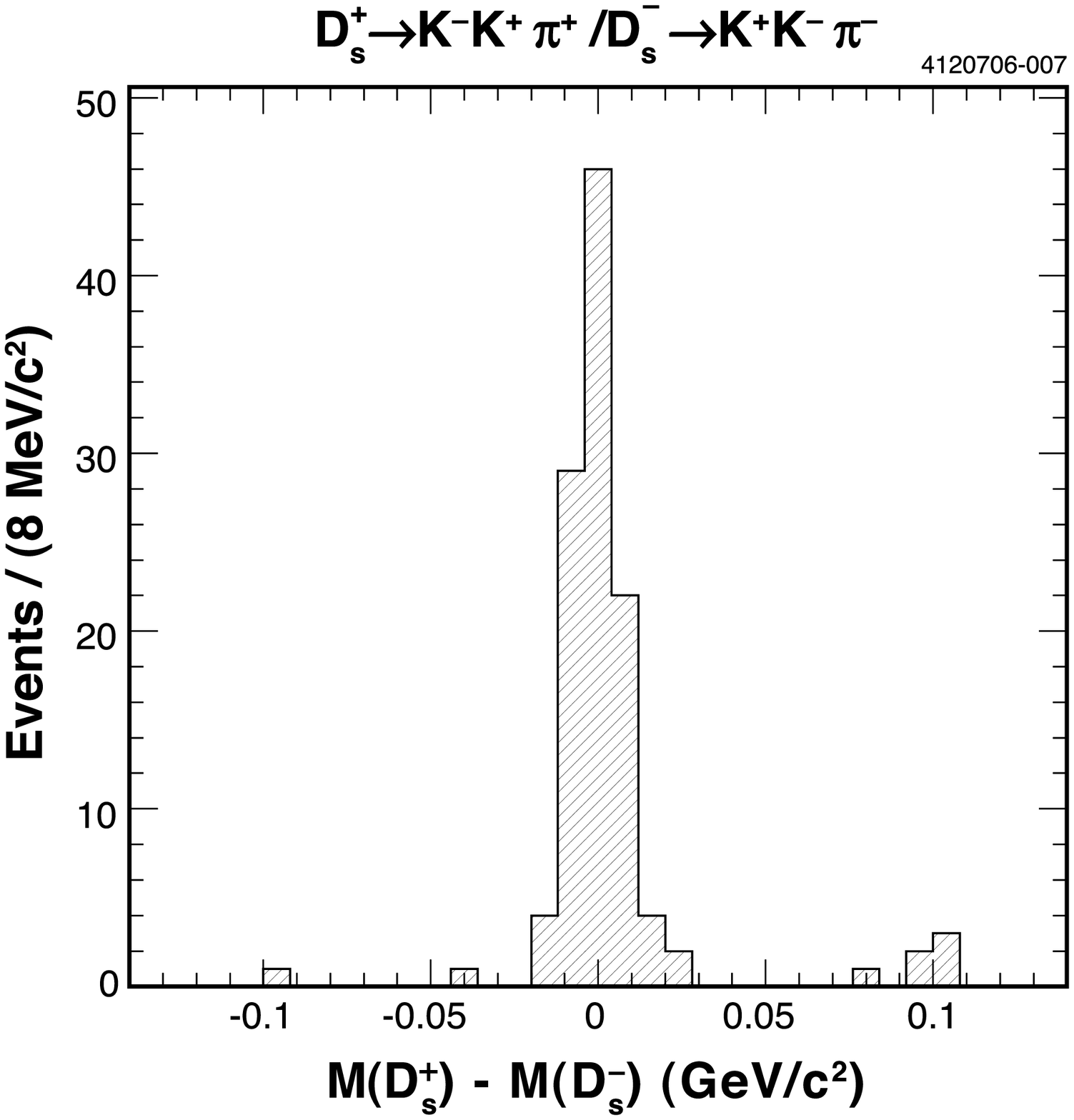}
\caption{Projections of  $\MDsm$ vs $\MDsp$ scatter plots on the $\DeltaM = \MDsp-\MDsm$ axis for  (left) all six $\Ds$ decay modes and (right) $\Dspm\to\Km\Kp\pipm$.}
\label{fig:dtproject}
\end{figure}

\clearpage

\def\etal{{\it et al.}}

\def\hepex#1{[\hbox{hep-ex/#1}]}

\end{document}